# Raman scattering investigation across the magnetic and MI transition in rare earth nickelate $R$NiO$_3$ ($R$ = Sm, Nd) thin films

C. Girardot[1,2], J. Kreisel[1,*], S. Pignard[1], N. Caillault[2], F. Weiss[1]

[1] Laboratoire Matériaux et Génie Physique, CNRS, Grenoble Institute of Technology, Minatec, 3, parvis Louis Néel, 38016 Grenoble, France

[2] Schneider Electric Industries S.A.S., 37 Quai Paul Louis Merlin, 38050 Grenoble Cedex 9 France

\* Corresponding author: *kreisel@inpg.fr*

**Abstract**

We report a temperature-dependent Raman scattering investigation of thin film rare earth nickelates SmNiO$_3$, NdNiO$_3$ and Sm$_{0.60}$Nd$_{0.40}$NiO$_3$, which present a metal-to-insulator (MI) transition at T$_{MI}$ and an antiferromagnetic-paramagnetic Néel transition at T$_N$. Our results provide evidence that all investigated samples present a structural phase transition at T$_{MI}$ but the Raman signature across $T_{MI}$ is significantly different for NdNiO$_3$ ($T_{MI} = T_N$) compared to SmNiO$_3$ and Sm$_{0.60}$Nd$_{0.40}$NiO$_3$ (T$_{MI}$ ≠ T$_N$). It is namely observed that the paramagnetic-insulator phase (T$_N$ < T < T$_{MI}$) in SmNiO$_3$ and Sm$_{0.60}$Nd$_{0.40}$NiO$_3$ is characterized by a pronounced softening of one particular phonon band around 420 cm$^{-1}$. This signature is unusual and points to an important and continuous change in the distortion of NiO$_6$ octahedra (thus the Ni-O bonding) which stabilizes upon cooling at the magnetic transition. The observed behaviour might well be a general feature for all rare earth nickelates with T$_{MI}$ ≠ T$_N$ and illustrates intriguing coupling mechanism in the T$_{MI}$ > T > T$_N$ regime.

I. Introduction

Rare earth nickelates with the generic formula $RE$NiO$_3$ ($RE$ = Rare Earth) have attracted a lot of research interest since the report of a sharp metal-to-insulator (MI) transition whereof the critical temperature $T_{MI}$ can be tuned with the rare earth size. Further to this, in the insulating phase most $RE$-nickelates exhibit a complex anti-ferromagnetic ordering below the Néel temperature $T_N$. Depending on the $RE$, these two transitions occur either at the same temperature or they are distinct.[1-4]

For a long time the crystal structure of $RE$-nickelates was thought to be orthorhombic (space group *Pbnm*) in both the metallic and the insulator regimes and the understanding of the *MI* transition remained somehow puzzling.[2,3] This view has been challenged by neutron diffraction and high-resolution synchrotron diffraction experiments which reveal two distinguishable NiO$_6$ octahedra in the insulator phase of $RE$NiO$_3$ leading to a monoclinic distortion with space group $P2_1/n$.[5-7] The observation of two independent Ni-positions has been interpreted by the presence of a long-range charge-disproparnation ($2\text{Ni}^{3+} \rightarrow \text{Ni}^{3+\delta} + \text{Ni}^{3-\delta}$) in the insulating regime, while nickel is uniformly trivalent in the metallic regime.[5] Recently, several authors[4,8,9] have proposed that the charge order in the insulating regime might induce improper ferroelectricity, thus multiferroicity, but this theoretical prediction is not yet experimentally verified. The charge-disproparnation scenario has been supported by a number of studies (e.g. ref. [10,11]) although other authors suggest a segregation into primarily ionic and primarily covalent bonding rather than charge transfer between the two Ni-sites.[12] Further to this, there is increasing evidence that $RE$-nickelates present local deviations from the average crystal structure. However, their electric, magnetic or structural nature is not yet clear and several scenarios like two-phase and bond fluctuations[12-14] or a short-range charge order even in the metallic state[15] have been suggested. It is usually assumed that both the low-temperature monoclinic charge disproparnation-type distortion and the short-range-order are shared by all members of the $RE$NiO$_3$ family and thus lead to a common mechanism of the MI-transition independently if $T_{MI} = T_N$ or $T_{MI} \neq T_N$.

In this paper two $RE$-nickelates have been investigated: SmNiO$_3$ and NdNiO$_3$. These two nickelates are particularly interesting to compare because for NdNiO$_3$ the magnetic and MI transitions occur according to literature at the same temperature ($T_{MI} = T_N \approx -80$ °C ) while they are distinct for SmNiO$_3$ ($T_{MI} \approx 125$ °C, $T_N \approx -100$ °C) and this although the ionic radii of Nd$^{3+}$ and Sm$^{3+}$ differ by only 0.03 Å. Furthermore, $T_{MI}$ can be readably adjusted[16] by varying $x$ in Sm$_{1-x}$Nd$_x$NiO$_3$ and our comparison has thus been extend by the composition Sm$_{0.60}$Nd$_{0.40}$NiO$_3$ which has a MI-transition near room temperature thus offering a promising suitability for applications.

The aim of our paper is threefold: (*i*) to verify if SmNiO$_3$ presents a symmetry breaking at the MI transition, knowing that such a symmetry breaking is not observed by high-resolution diffraction, (*ii*) to investigate if the mechanisms at the MI transition is driven by the same structural phase transition at $T_{MI}$ for $T_{MI} = T_N$ and $T_{MI} \neq T_N$ systems, (*iii*) to investigate if the phonon signature is notably affected by the magnetic and electric transitions.

In order to address the above questions, we have used Raman spectroscopy which is known to be a versatile technique for the investigation of thin film oxides and in particular for the detection of even subtle structural distortions in perovskites.[17-21] Phonons are also known to be influenced by spin and electronic correlation thus offering a complementary tool for understanding transition metal oxides.

**II – Experimental and sample characterisation**

The preparation of $RE$NiO$_3$ materials in bulk form is difficult due to the necessity of an important oxygen pressure to stabilize Ni in its 3+ state of oxidation.[1] A way to circumvent this difficulty is to stabilize $RE$NiO$_3$ by epitaxial strain in thin film growth.[16, 19, 22] Most of the used deposition techniques, such as pulsed laser deposition, also need high oxygen pressure in order to fabricate the targets. On the other hand, a low-pressure Metal Organic Chemical Vapour Deposition (MOCVD) process offers an opportunity[22] to obtain $RE$NiO$_3$ phases without using high oxygen pressures. Here, we have used an 1:1 Ar-O$_2$ ratio with a flux of 600 cm$^3$/min. Nevertheless, $RE$NiO$_3$ can be only synthesized by MOCVD on cubic or pseudo-cubic single crystalline substrates presenting a lattice parameter close to that of the nickelate. It is then the stress imposed by the substrate to the film which leads to the thermodynamic stabilization of the nickelate phase.

Our thin films were obtained by injection Metal-Organic Chemical Vapour Deposition (MOCVD) "band flash" using 2,2,6,6-tetramethylheptanedionato-chelates of corresponding metals as volatile precursors. Polished single crystalline substrates of perovskite-type (00l) LaAlO$_3$ (LAO) was used to achieve an epitaxial stabilisation of the films. Deposition conditions were 680°C using an atmosphere of argon-oxygen at 10 mbar pressure, followed by an in-situ annealing during 30 minutes at ambient pressure under a pure oxygen flow. It has been previously reported that this technique and conditions lead to good quality $RE$-nickelate thin films on various substrates.[23-26]

The thickness and chemical composition of the films were determined by Wavelength Dispersive Spectroscopy (WDS) using a CAMECA SX50 spectrometer. A slight excess of Ni is measured corresponding to a minor NiO (220) oriented phase which is detected by X-ray diffraction on a Siemens diffractometer in a θ/2θ geometry with a Cu anode ($\lambda$ =1.5406 Å). Furthermore, X-ray diffraction indicates that the SmNiO$_3$ (SNO), Sm$_{0.60}$Nd$_{0.40}$NiO$_3$ (SNNO) and NdNiO$_3$ (NNO) films are textured on the (00l) oriented LaAlO$_3$ substrate. Thicknesses are respectively 17 ± 3 nm, 52 ± 5 nm and 75 ± 5 nm. A transmission Electronic Microscopy study of the SNO film (not shown) testifies an atomically flat and coherent interface.

The temperature-dependent electrical characterisation of the thin films was measured from - 270 to 250 °C by the well-known four-probe technique[27] using a DC current and platinum wires with a conducting silver paint (Ag-epoxy, each separated by 0.5 mm). The surface of the contacts is about 0.5 mm². The value of the measured resistance is obtained by the average of four measurements at different currents injected by a precision current source Keitley 6220 on the external electrodes. A nanovoltmeter Keitley 2182A was used to measure the voltages across the electrical leads, with two internal probes ($V_{meas}$), to obtain the resistance of the sample. The resistivity $\rho$ is calculated from the resistance $R$ by using equation (1):

$$\rho = \frac{\pi . w}{\ln 2} R$$

with $w$, being the thickness of the film.[28] Figure 1 shows the temperature-induced evolution of the resistivity for different rare earth nickelate thin films and illustrates that all three films undergo a MI-transition et $T_{MI}$ and that $T_{MI}$ depends critically on the size of the rare earth. The observed $T_{MI}$'s are determined from the point of deflection of the resistivity curve at 120 °C (SNO), 45 °C (SNNO) and -115 °C (NNO, on heating). These values differ from the values reported in literature on bulk samples, the difference can be attributed to the effect of strain in the thin films. In agreement with literature, the resistivity curve of NNO presents a hysteresis indicating a first-order phase transition; such a hysteresis is not observed for SNO and SNNO.

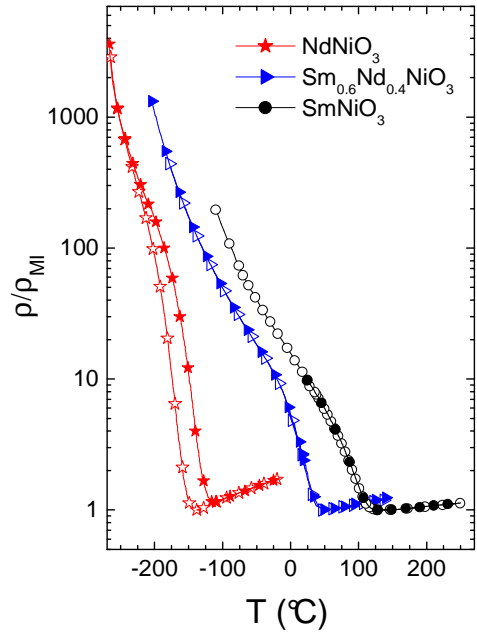

*Figure 1 (colour on-line)*

*Resistivity vs. temperature for different rare earth nickelate thin films [SmNiO$_3$ (SNO, ●/○), Sm$_{0.60}$Nd$_{0.40}$NiO$_3$ (SNNO, ▶/▷ ) and NdNiO$_3$ (NNO, ★/☆)]. The full and open signals correspond to heating and cooling, respectively. $\rho_{MI}$ is the resistivity at $T_{MI}$.*

Raman spectra were recorded with a LabRam Jobin-Yvon spectrometer. The 514.5-nm line of an Ar$^+$ ion laser was used as excitation line. Experiments were conducted in micro-Raman; the light was focused to a 1μm$^2$ spot. All measurements performed under the microscope were recorded in a back-scattering geometry; the instrumental resolution was 2.8 ± 0.2 cm$^{-1}$. It has been reported earlier that Raman spectra recorded on black manganite[21, 29] and nickelate thin films[19] can show a strong dependence on the exciting laser power. As a consequence, standard experiments have been carried out using low powers (less then 1 mW under the microscope, using a times 50lf objective) and it has been verified that no structural transformations and overheating takes place. Namely, the below discussed difference in $T_{MI}$ between Raman scattering and conductivity measurements and between bulk and thin film data does not result from laser heating. Temperature-dependent Raman measurements have been carried out by using a commercial LINKAM heating stage placed under the Raman microscope. After an initial cooling, all presented spectra have been obtained by heating the sample from low-temperature (-180 °C) above room temperature (up to 150 °C), this is in the following simplified as "heating". The Raman spectra before and after temperature measurements are identical, attesting the reversibility of temperature-induced changes.

### III – Raman scattering of $Sm_{1-x}Nd_xNiO_3$ thin films

#### A. General considerations

The high-temperature metallic phase of *RE*-Nickelates is an orthorhombically distorted perovskite with space group *Pbnm*. With respect to the ideal cubic *Pm*3*m* perovskite structure the orthorhombic structure is obtained by an anti-phase tilt of the adjacent $NiO_6$ octahedra ($a^-a^-c^+$ in Glazer's notation[30]). *RE*-Nickelates are thus ferroelastic. The 10 atoms in the unit cell of the orthorhombic structure give rise to 24 Raman-active modes[19]

$$\Gamma_{Raman, Pbnm(metallic)} = 7A_{1g} + 7B_{1g} + 5B_{2g} + 5B_{3g}$$

The proposed[5-7] low-temperature monoclinic structure of *RE*-Nickelates (space group $P2_1/n$) presents also 24 Raman-active modes.

$$\Gamma_{Raman, P2_1/n(insulator)} = 12A_g + 12B_g$$

As a consequence of this, the number of Raman bands alone does not allow distinguishing between the orthorhombic and monoclinic space group, but changes in the spectral signature are expected to mirror any change in symmetry. To date Raman scattering data on *RE*-nickelates remain still limited in the literature and the authors are only aware of a temperature-dependent investigation of $NdNiO_3$ (ref. [19]) and a room temperature spectrum of $SmNiO_3$ (ref. [23]).

**B. Room-temperature spectra**

Let us first remind that NNO is at room temperature metallic (orthorhombic), SNO is insulating (expected monoclinic) and SNNO is close to the metal-insulator transition. Figure 2 presents a comparison of room temperature Raman spectra for SNO, SNNO and NNO. The spectrum of NNO is similar to the data reported in the literature[19] and clearly distinct from the SNO spectrum, suggesting that NNO and SNO do not adopt the same crystal structure at room temperature. It can also be seen that the Raman signature of SNNO is rather similar to that of NNO which in turn suggest that SNNO is orthorhombic at room temperature.

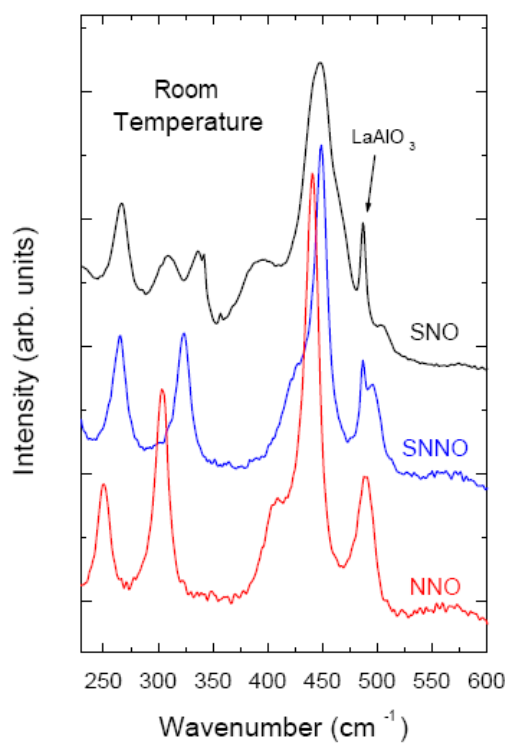

*Figure 2 (colour on-line)*

*Comparison of room temperature Raman spectra for $SmNiO_3$ (SNO), $Sm_{0.60}Nd_{0.40}NiO_3$ (SNNO) and $NdNiO_3$ (NNO). The arrows indicate Raman bands from the $LaAlO_3$ substrate.*

## C. Temperature-dependent Raman scattering

Figure 3 presents temperature-dependent Raman spectra for SNO, NNO and SNNO while Figure 4 shows how the band positions change with temperature for SNO and SNNO.

The spectra of NNO are similar to those previously reported in ref. [19]. Although the low-wavenumber soft mode is not observed due to the Notch-filter cut-off of our spectrometer a phase transition around $T_{MI} = T_N = -115$ °C is evidenced by significant changes in the spectral signature: When going from low- to high-temperature one notices namely (*i*) the change from of one strong band ($\approx 300$ cm$^{-1}$) and one weak band ($\approx 330$ cm$^{-1}$) towards a new single band around 310 cm$^{-1}$, (*ii*) a sudden low-wavenumber shift of some of the bands around 420 cm$^{-1}$ and (*iii*) a disappearance of the bands around 400 cm$^{-1}$.

All these features of Figure 3.c are clear evidence for a structural phase transition of NNO at $T_{MI}$. Raman scattering on its own does not allow the determination of the two space groups but it is likely that the signature corresponds to the transition proposed in literature[10, 11, 19] when heating NNO: monoclinic *P*2$_1$/*n* → orthorhombic *Pbnm*. We further note that a phase coexistence is observed at $T_{MI}$ as namely evidenced by the presence of both the "monoclinic" band at 300 cm$^{-1}$ and the "orthorhombic" band at 310 cm$^{-1}$ at the critical temperature. This is in agreement with, and further evidence of, a first-order phase transition of NNO at $T_{MI}$. When the temperature is further increased in the metallic phase up to room temperature the Raman signature presents no changes apart from a slight temperature shift due to thermal expansion and thermal broadening of all bands.

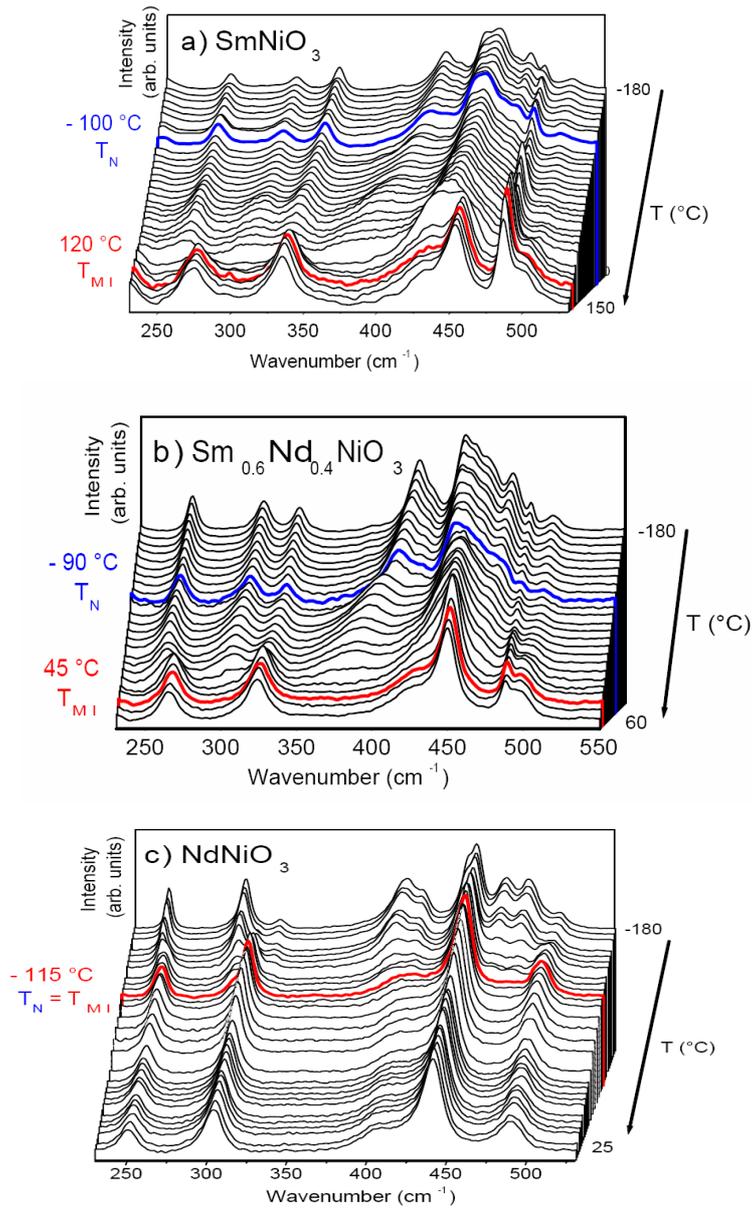

*Figure 3 (colour on-line)*

Selected temperature-dependent Raman spectra for SmNiO$_3$ (SNO, top), Sm$_{0.60}$Nd$_{0.40}$NiO$_3$ (SNNO, middle) and NdNiO$_3$ (NNO; bottom). The displayed spectra have been obtained by heating from -180°C with a step of 5°C, for simplicity only every second spectrum is shown. The red spectra correspond to MI-transition temperatures (deduced from Figure 1), while the blue spectra correspond to magnetic Néel-transition temperatures (bulk literature data). Note that $T_{MI}$ measured by the electrical four-probe technique (in red) is by roughly 20 °C higher than the observed anomalies in the Raman spectra (see text).

We will now discuss the temperature-dependent signature of SNO. At a low-temperature of -180 °C SNO is an antiferromagnetic insulator and the Raman signature of SNO is similar to that of NNO, which suggests that SNO and NNO adopt the same crystal structure. Since bulk NNO is reported[19] to be monoclinic at low-temperature this similarity suggests that SNO also adopts a monoclinic structure in the low-temperature range below $T_{MI}$. Upon heating SNO, enters at $T_N$ the paramagnetic phase where it shows first significant changes in its Raman spectra: It can be seen in Figure 3.a and Figure 4 that the band at initially 420 cm$^{-1}$ shows from $T_N$ to $T_{MI}$ a significant softening of 30 cm$^{-1}$ towards lower wavenumbers. Furthermore, a decrease in intensity (relative to the massif at 450 cm$^{-1}$) and an increase in its width of the same band are observed. Although this softening is the predominant signature of the intermediate paramagnetic insulator phase other less visible spectral changes like the progressive intensity loss of the band at 320 cm$^{-1}$ (relative to the massif at 450 cm$^{-1}$) accompany the spectral evolution. The possible origins of the softening are discussed in more detail in the discussion section IV of this manuscript. When the temperature is now further increased SNO enters at $T_{MI}$ the paramagnetic-metallic phase where new significant changes in the spectral signature are observed, three changes are predominant (Figures 3 and 4): (*i*) the suppression of the above-described 420 cm$^{-1}$ mode and the band at initially 320 cm$^{-1}$ (*ii*) an abrupt change in wavenumber of the band at initially 270 cm$^{-1}$, (*iii*) a significant change in the intensity distribution of the massif between 400 and 450 cm$^{-1}$. Again all these spectral changes offer direct evidence for a structural phase transition at $T_{MI}$, which most likely corresponds to a transition from a low-temperature monoclinic phase to the well-known *Pbnm* orthorhombic phase above $T_{MI}$.

Although the electrical and Raman measurements are done on physically the same thin film, the $T_{MI}$ measured by Raman scattering is roughly 20 °C lower than what is observed with the electrical four-probe technique. This observation is likely due to the fact that Raman scattering is a local probe while the electric measurement probes a macroscopic phase transition. In agreement with literature, this apparent discrepancy between electrical and Raman measurements offers thus further hints for local structural/electric deviations from the average structure in *RE*-nickelates.

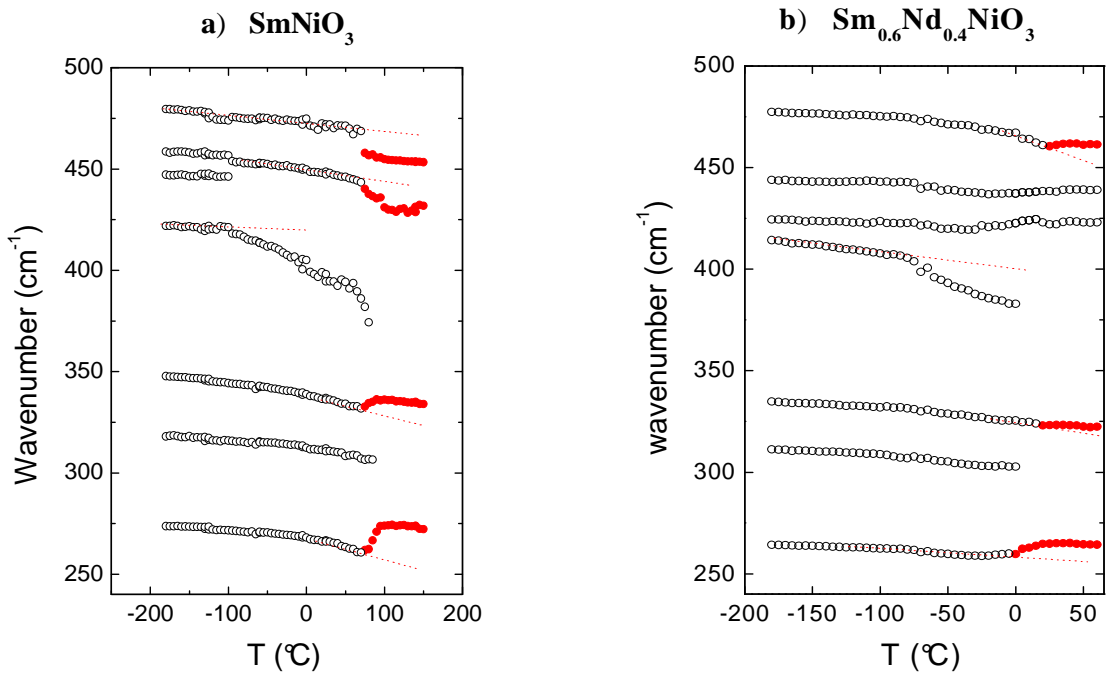

*Figure 4 (colour on-line)*

*Temperature-dependent evolution of the band position in the Raman spectra of a) $SmNiO_3$ and b) $Sm_{0.60}Nd_{0.40}NiO_3$. Lines and colours are guides to the eye to emphasize spectral changes which in turn are evidence for a structural phase transition at $T_{MI}$ (see text).*

Having discussed the spectral signatures of $SmNiO_3$ and $NdNiO_3$, it is now interesting to inspect the signature of a member of the solid solution $Sm_{0.60}Nd_{0.40}NiO_3$ (SNNO). It can be seen from Figure 3.b that the overall evolution of the Raman spectra of SNNO with temperature is very similar to SNO but different from NNO. Note that the temperature range of the intermediate paramagnetic insulator phase ($T_N < T < T_{MI}$), by using the range of the mode softening as an indicator, extends over a smaller temperature range for SNNO when compared to SNO, in agreement with earlier results from electric and magnetic measurements.[2, 16]

**IV – Discussion**

It has been shown in the previous section that the temperature-dependent Raman signature through the MI transition is different for a *RE*-Nickelate where $T_{MI} = T_N$ (NNO) compared to those where $T_{MI} \neq T_N$ (SNO, SNNO). The difference is revealed by the observation that the paramagnetic insulator phase (which only exists in the $T_{MI} \neq T_N$ systems) has a specific Raman signature: The predominant signature is a mode at $\approx 420$ cm$^{-1}$, which softens significantly to lower wavenumbers when the antiferromagnetic-paramagnetic phase transition at $T_N$ is crossed upon heating, leading to an important shift of 30-40 cm$^{-1}$ between $T_N$ and $T_{MI}$, thus in the paramagnetic insulating phase (Figure 5). This pronounced softening indicates an important anharmonicity which cannot be explained by a classical temperature behavior but points at a further contribution. A straightforward interpretation of this softening is difficult because a number of different explanations might be considered. The aim of the following part will be to illustrate different possible origins and to cut the discussion down to the most likely scenarios.

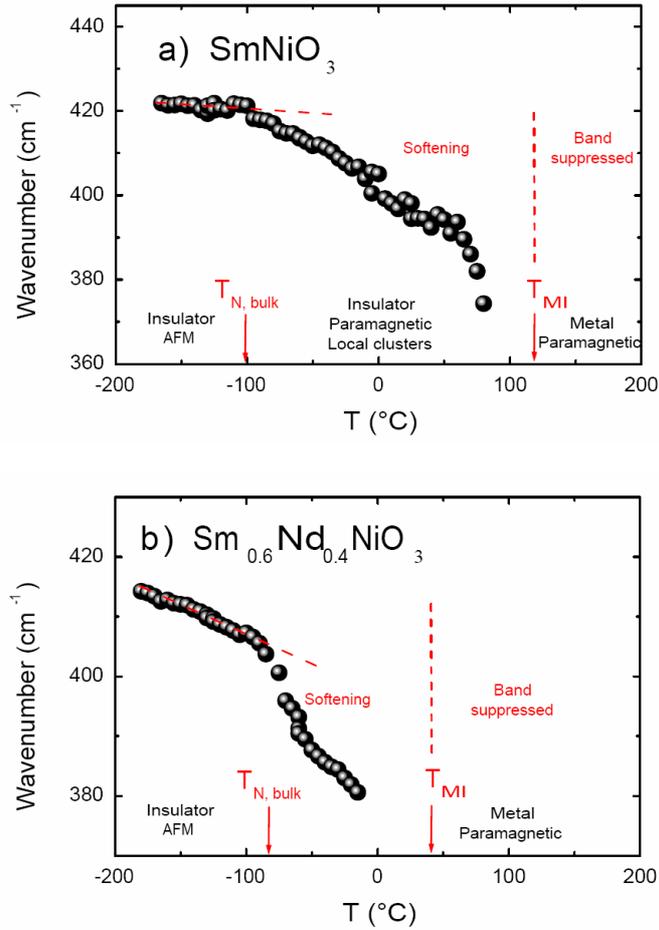

*Figure 5 (colour on-line)*

*Temperature-dependent evolution of the band showing a significant softening for a) $SmNiO_3$ and b) $Sm_{0.60}Nd_{0.40}NiO_3$. Lines and colours are guides to the eye to underline spectral changes at the magnetic ($T_N$) and electronic ($T_{MI}$) phase transitions, see text for more details. $T_N$ corresponds to literature results on bulk samples, while $T_{MI}$ corresponds to the measured metal-insulator phase transition on physically the same films which have been investigated by Raman scattering.*

At first sight, the considerable spectral changes at $T_N$ for SNO and SNNO could be an indication of a structural rearrangement which has to be subtle or to occur on a local level since it is not detected by X-ray or neutron diffraction. Although we cannot formally exclude a change in the space group at $T_N$, such a scenario is very unlikely due the absence of a new spectral signature in terms of new bands, band splitting etc. Given the fact that the softening starts at $T_N$, one should consider magnetostriction. However, the usually observed changes in lattice parameter at $T_N$ are by far too small to cause the observed phonon anomaly, especially over a broad range of temperature.

It is useful to recall that the important softening is mainly observed for one band, which indicates that a particular type of chemical bonding (and its associated force constant) change with temperature. In perovskites, the spectral region of this band is usually associated with vibrations of the octahedron and, in agreement with literature[19], we thus attribute the 420 cm$^{-1}$ band to a mode involving distortions of the NiO$_6$ octahedra which in turn is a signature for significant changes in the Ni-O bonding. This brings us back to the fact that it has been proposed that the Ni-O bonding is considerably modified at $T_{MI}$, whereof one proposed mechanisms is the following charge-disproportation upon cooling:

$$2Ni^{3+} \text{ (metallic, orthorhombic)} \rightarrow Ni^{3+\delta} + Ni^{3-\delta} \text{ (insulating, monoclinic)}$$

Such a mechanism conditions on a local level a significant change in chemical bonding since the charge dispropornation leads to an increased covalent character of the Ni-O bond. The charge order, i.e. the magnitude of $\delta$, has been earlier proposed[31] to be the order parameter close to the MI-transition. Within this picture it appears natural to consider that the softening of the band at 420 cm$^{-1}$ might be the signature of a temperature-dependent increase of the charge disproporation, i.e. $\delta$ increases with decreasing temperature up to the magnetic phase transition from whereof it is rather stabilised. However, to the best of our knowledge, there is no example in the literature where a change in a charge dispropornation alone has conditioned a large phonon softening.

Finally, a possible coupling between spin and phonon degrees of freedom can be considered. A useful approach for the understanding of spin-dependent phonon frequencies is based on an initial model by Baltensperger and Helmann [32] where it has been proposed that the phonon frequencies in magnetic materials are affected by the correlation of nearest-neighbour spin pairs.[32, 33] Such a scenario should be considered to explain also the observed anomaly in SNO and SNNO but the amplitude of the here reported softening remains large compared to what is known for other magnetic materials, : e.g. the structurally similar orthoferrite $EuFeO_3$ presents at its AFM-PM phase transitions a phonon anomaly of an order of magnitude smaller than what we observe for SNO.[34] The spin-phonon coupling in other materials such as $MF_2$ ($M$ = Fe, Mn)[33], $Y_2Ru_2O_7$ (ref. [35]), $LaTiO_3$ (ref. [36]) or $ZnCr_2O_4$ (ref. [37]) is also significantly smaller than in our observation.

Nevertheless, we believe that the observation of the anomaly around $T_N$ suggests that spin-phonon coupling is a contributing mechanism but it is unlikely that spin-phonon coupling alone can explain the observed softening. We thus consider a coupling to another order parameter which does change with temperature. One common order parameter in perovskites is cation displacement thus ferroelectricity. As mentioned in the introduction, recent reports propose that *RE*-Nickelates present in their insulating a charge-order driven (thus improper) ferroelectricity due to the polar displacements of Ni-cations.[4, 8, 9] Other authors discuss the possibility of potential ferroelectricity in the antiferromagnetic *E*-Phase of the orthorhombic ground state of nickelates.[8, 38] Ferroelectric soft modes usually occur at lower wavenumbers. On the other hand, the remarkable softening observed in SNO and SNNO might be related to the fact that both magnetic and ferroelectric order coexist and are temperature-dependent at the same time. This allows drawing the following tentative scenario: The onset of magnetic correlations takes place within a temperature regime that presents ferroelectric lattice instabilities, i.e. in a regime where the Ni-cations are easily displaced in respond to any perturbation (here magnetic correlations). This scenario is consistent with our assignment of the softening band to distortions of the $NiO_6$ octahedra (thus Ni-O bonding). This scenario calls for two notes: First, our observed softening extends deep into the paramagnetic phase thus it has to be primarily driven by *local* magnetic correlations; as stated in the introduction such a local (cluster-like) structure is commonly accepted for *RE*-Nickelates.[12-15] Second, the local structure is namely characterized by changes in the charge order (in magnitude and size) which in turn drives the hypothetical ferroelectricity. As a consequence, one can expect an intimate correlation between magnetism, charge order and ferroelectricity in this highly correlated electron system. The above-described coupling, however, is a hypothesis which remains to be elucidated namely by ferroelectric measurements which are unfortunately (further to conductivity problems) not straightforward for films on an insulating substrate as is the case for the here investigated samples.

**V - Conclusion**

$SmNiO_3$, $NdNiO_3$ and $Sm_{0.60}Nd_{0.40}NiO_3$ thin films have been investigated by temperature-dependent Raman scattering across their magnetic and *MI* transitions. The spectral signature provides evidence that all investigated samples present a structural phase transition at $T_{MI}$. Interestingly, the Raman signature across $T_{MI}$ is significantly different for $NdNiO_3$ ($T_{MI} = T_N$) compared to $SmNiO_3$ and $Sm_{0.60}Nd_{0.40}NiO_3$ ($T_{MI} \neq T_N$), thus suggesting that the mechanism are not the same. It has been observed that the paramagnetic-insulator phase ($T_N < T < T_{MI}$) in $SmNiO_3$ and $Sm_{0.60}Nd_{0.40}NiO_3$ is characterized by a pronounced softening of one particular phonon band around 420 cm$^{-1}$. This signature is unusual and points to important and continuous modifications in Ni-O bonding, which stabilize upon cooling at the magnetic transition. The observed behaviour might well be a general feature for all rare earth nickelates with $T_{MI} \neq T_N$.

We relate this softening to spin-phonon coupling and hypothesize that a coupling to a ferroelectric order parameter conditions the unusual strength of the softening. In this scenario the softening is thus directly related to the theoretically[4, 8, 9] predicted magnetoelectric multiferroic character of the insulating regime in SNO and SNNO and would be the first experimental hint that *RE*-Nickelates are indeed multiferroic. More experimental and theoretical work is needed, to unambiguously identify the physical mechanisms leading to the anomalies reported here and to confirm the multiferroic character. First principle calculations of the phonon spectrum with models including magnetic correlations, the peculiar local structure as well as lattice instabilities could lead to a better understanding of the proposed coupling.

*Acknowledgements*

This work was supported by the European Network of Excellence FAME (Functionalized Advanced Materials and Engineering), by Schneider Electric S.A and the European Strep MaCoMuFi. The authors thank H. Roussel (from the "Consortium des Moyens Technologiques Communs", CMTC) for x-ray characterization and A. Pasturel (SIMAP, Grenoble) for enlightening discussion on metal-insulator transitions. J. Marcus (Institut Néel, Grenoble) is acknowledged for access to electric transport measurements.